\documentclass[]{interact}
\usepackage{hyperref}

\usepackage{epstopdf}% To incorporate .eps illustrations using PDFLaTeX, etc.
\usepackage[caption=false]{subfig}% Support for small, `sub' figures and tables

\usepackage[longnamesfirst,sort]{natbib}% Citation support using natbib.sty
\bibpunct[, ]{(}{)}{;}{a}{,}{,}% Citation support using natbib.sty
\renewcommand\bibfont{\fontsize{10}{12}\selectfont}% To set the list of references in 10 point font using natbib.sty

\theoremstyle{plain}% Theorem-like structures provided by amsthm.sty

\theoremstyle{definition}

\theoremstyle{remark}

\begin{document}

%\articletype{ARTICLE TEMPLATE}% Specify the article type or omit as appropriate

\title{An Empirical Evaluation of Text Representation Schemes on Multilingual Social Web to Filter the Textual Aggression}

\author{
\name{Sandip Modha \textsuperscript{a}\thanks{CONTACT Sandip Modha. Email: sjmodha@gmail.com} and Prasenjit Majumder\textsuperscript{a} \thanks{CONTACT Prasenjit Majumder. Email: prasenjit.majumder@gmail.com}}
\affil{\textsuperscript{a}DA-IICT Gandhinagar,India; }
}

\maketitle
\date{}
\begin{abstract}
Due to an exponential rise in the social media user base, incidents like hate speech, trolling, cyberbullying are also increasing and that lead hate speech detection problem reshaped into different tasks like Aggression detection, Fact detection. This paper attempt to study the effectiveness of text representation schemes on two tasks namely: User Aggression and Fact Detection from the social media contents. In User Aggression detection, The aim is to identify the level of aggression from the contents generated in the Social media and written in the English, Devanagari Hindi and Romanized Hindi. Aggression levels are categorized into three predefined classes namely: `Non-aggressive`, `Overtly Aggressive`, and `Covertly Aggressive`. During the disaster-related incident, Social media like, Twitter is flooded with millions of posts. In such emergency situations, identification of factual posts is important for organizations involved in the relief operation. We anticipated this problem as a combination of classification and Ranking problem. This paper presents a comparison of various text representation scheme based on BoW techniques, distributed word/sentence representation, transfer learning on classifiers. Weighted $F_1$ score is used as a primary evaluation metric. Results show that text representation using BoW performs better than word embedding on machine learning classifiers. While pre-trained Word embedding techniques perform better on classifiers based on deep neural net. Recent transfer learning model like ELMO, ULMFiT are fine-tuned for the Aggression classification task. However, results are not at par with pre-trained word embedding model. Overall, word embedding using fastText produce best weighted $F_1$-score than Word2Vec and Glove. Results are further improved using pre-trained vector model. Statistical significance tests are employed to ensure the significance of the classification results. In the case of lexically different test Dataset, other than training Dataset, deep neural models are more robust and perform substantially better than machine learning classifiers. 

\end{abstract}

\begin{abbreviations}
NLP: Natural Language Processing; BoW :Bag-of-Word; CNN:Convolution Neural Network;LSTM :Long Short-Term Memory.
\end{abbreviations}

\section{Introduction}

\label{intro}
The Social Web is a great source for studying human interaction and behavior. In the last few years, there is an exponential growth in Social Media user base. Sensing content of Social Media like Facebook, Twitter, by the smart autonomous application empower its user community with real-time information which is unfolded across the different part of the world. Social media provide the easiest and anonymous platform for common people to voice their opinion or view on a various entity like celebrity, politician, product, stock market etc or any social movement. Sometime such opinions might be aggressive in nature and propagate hate in the social media community. 

With the unprecedented increase in the user base of the social media and its availability on the Smartphones, incidents like Hate speech, trolling, Cyberbullying, and Aggressive posts are increasing exponentially. A smart autonomous system is required which enable surveillance on the social media platform and detect such incidents. Some of the researchers look posts from the aspect like aggression \citet{Kumar18} to filter the contents. some of the posts contain words which might be qualified as either highly or overly aggressive or have hidden aggression. Sometimes posts do not have any aggression. Based on these, posts or comments are categorized into three classes namely: `Overtly Aggressive`, `Covertly Aggressive` and `Non-aggressive`  \citet{Kumar18}. Henceforth, in the rest of the paper, we will denote these classes by these abbreviations namely: OAG, CAG, NAG respectively.Table ~\ref{tab:Sample-post} shows the sample posts belonging to these classes.

\begin{table}
\tbl{sample post for the each class.}
{\begin{tabular}{p{0.6cm} p{8cm}p{2cm}} \toprule
 
   &Post text & Class Label \\ \midrule
1 & She is simple girl and need not know politics. Let her vote to her choice & NAG \\

2 &People talk about common man suffering all the time.This is the same country where thousands laid their lives for freedom.Cant the common man endure little trouble to stand in queues for the greater good
 & CAG \\

3 & Langove get out from this conversation. U are an uninvited dog here  & OAG\\ \bottomrule

\end{tabular}}
\label{tab:Sample-post}
\end{table}

Social Media, specifically Microblog has proved its importance during the disaster-related incidents like an earthquake, Hurricane and floods \footnote{https://phys.org/news/2018-08-social-media-bad-disaster-zones.html}. Organizations involved in relief operation actively track posts related to situational information posted on Facebook and Twitter during the disaster. However, At the same time, social media is flooded with lots of prayer and condolence messages. Posts which contain factual information are extremely important for the organization involved in post-disaster relief operations for coordination. Filtering and Ranking of the posts containing factual information will be very useful to them. We believe that this is the special problem of the Sentiment Analysis task. We consider this problem as a combination of two-class classification problem: factual posts and nob-factual posts plus Ranking. Table \ref{tab:Sample-post-so} shows  the example of the posts of belong to these class.

\begin{table}
\tbl{sample post for the each class.}
{\begin{tabular}{p{0.6cm} p{8cm}p{2cm}} \toprule
 
   &Post text & Class Label \\ \midrule
1 & \#Nepal \#Earthquake day four. Slowly in the capital valley Internet and electricity beeing restored . A relief for at least some ones & Factual \\

2 & PMOIndia Indian Government is doing every possible help to the earthquake victims and they need money so plz contribute & Non-factual \\ \bottomrule

\end{tabular}}
\label{tab:Sample-post-so}
\end{table}

The Text representation of social web content plays a pivotal role in any NLP task. Bag-of-word is the oldest and simple technique to represent the document or post into a fixed length vector. The BoW techniques generate very sparse and high dimensional space vector. Text representation using distributed word/sentence representation or word embedding is gain rapid momentum recently. In this paper, one of the objectives is to find the best text representation scheme to model social web content for the machine learning classifier and deep neural net. Various Text representation scheme based on BoW, word embedding and are studied empirically. We have reported result on popular word embedding technique like Word2vec, Glove and fastText on standard machine learning classifier like Multinomial Naive Bayes (MNB), Logistic Regression(LR), K-Nearest neighbors KNN, Support Vector Classifier (SVC), Decision Tree (DT),Stochastic Gradient Descent(SGD), Random forest (RF), Ridge, AdaBoost, Perceptron, Deep neural net based on LSTM, CNN and Bidirectional LSTM. Results  are also reported on Doc2vec embedding, a popular sentence or paragraph embedding technique for the above classifiers.

Transfer Learning is well practiced in the area of computer vision. However, in the NLP, transfer learning has limited application in the form of pre-trained word vector which is used to initialize the weights of the embedding layer of the deep neural network. With the advent of transfer learning method like ELMO \citep{peters2018deep}, ULMFiT \citep{howard2018universal} claimed substantial improvement in the performance of various NLP tasks like Sentiment Analysis, Question/Answering, Textual Entailment empirically. The main idea behind these methods is to train language model on the large corpus and fine tune on the task-specific corpus. In this paper, We have evaluated the performance of these methods in the Aggression classification tasks.  

\subsection{Research Questions}
In this study, experiments are performed on the benchmark dataset with to answer the following questions
\begin{itemize}
\item Which is the best Text Representation scheme to model text from the Social Web?
\item  Does pre-trained language model based on transfer learning better than pre-trained word embedding based on shallow transfer learning on Social media data? 
\item Does Making too Deep Neural net make sense?
\end{itemize}

To answer all research question listed above, experiments are performed on two tasks namely: Aggression detection (Trolling Aggression and Cyberbullying (TRAC) dataset) \cite{trac2018dataset} and Fact detection (FIRE iRMDI Dataset)\cite{fire2018-irdimis}. In this paper, we present exhaustive benchmarking of text representation schemes on these datasets. Our results reveal that fastText with pre-trained vector along with CNN outperform standard machine learning classifiers based on BoW Model and  marginally perform better than Word2vec and Glove. Paragraph vector or Doc2vec \cite{le2014distributed} perform very poor on our dataset and turn out to be the worst text representation scheme among all. We also found that model based on the deep neural net is more robust than machine learning classifier when tested on lexically different dataset than training Dataset. i.e. deep neural model substantially outperforms machine learning classifier on Twitter test Dataset while trained on Facebook Dataset in this evaluation. 

To validate our claims, statistical significance tests are performed on weighted $F_1$-score of the classifier for each text representing scheme. Statistical inference is used to check evidence to support or reject these claims. Significance tests like Wilcoxon signed-rank and Student t-test were carried out by comparing  weighted $F_1$ score all the text representation scheme with the fastText pre-trained vector. In most of the cases, p-values are less than 0.05.  

The rest of the paper is organized as follows: In section \ref{Related-Work}, we review the relevant works in the area of Sentiment analysis and hate speech detection.  Section \ref{Dataset} contains the detail information about the various benchmark Datasets used in the experiments. Various Text Representation schemes are described in section \ref{Text-rep-scheme}. We formally describe the evaluation task and models in section \ref{task}. We report results in section \ref{sec:results} and present detail result analysis in section \ref{lab:ra}. We conclude the discussion and provide insight for the future work in section \ref{lab:con}.

\section{Related Work}
\label{Related-Work}
Bag-of-Words (BoW) \citep{harris1954distributional} is the oldest technique to represent the text of the documents in fixed-length vectors with high dimensionality. \citet{mikolov2013efficient} proposed two architecture namely: skip-gram(SG) and continuous-bag-of word (CBOW) to learn high quality low dimensional word embedding. However, to generate sentence vector often, average or mean of word vector are considered. Doc2vec or paragraph vector \citet{le2014distributed} proposed Paragraph2vec (Doc2vec)  which is the extension of the Word2vec to learning document level embedding. It is an unsupervised method which learns document vector from paragraph, sentence or document. \citet{pennington2014glove} proposed word embedding based on the co-occurrence matrix. \citet{lau2016empirical} have performed a comprehensive evaluation of Doc2Vec on two tasks namely: Forum Question Duplication and Semantic Textual Similarity (STS) task. Authors claimed that Doc2Vec performs better than Word2vec provided that models trained on large external corpora, and can be further improved by using pre-trained word embedding. They have published the hyper-parameter for the Doc2Vec embedding. Our work is similar to this but we have reported the evaluation of all the text representation scheme including doc2vec on TRAC dataset\citep{trac2018dataset} on each classifier.

Hate speech is a type of language which is used to incite or spread violence towards the group of people based on the gender, community, race, religion. Sentiment analysis and hate-speech are closely related in fact sentiment analysis techniques are used in hate speech detection. Initially, Sentiment Analysis problem is formulated as a binary classification problem for predicting the election results or detecting political opinion \citep{maynard2011automatic, conover2011political, conover2011predicting, tumasjan2010predicting} on Twitter. Then after, It turned into the multi-class classification problem with the introduction of the neutral label. Soon, Researchers come with different notion like aggression \citep{Kumar18}, cyberbullying\citep{xu2012learning}, sarcasm, trolling. Semeval (International workshop on semantic evaluation)\citep{rosenthal2017semeval} is one of the popular competition on sentiment analysis which is started since 2013. TRAC \footnote{https://sites.google.com/view/trac1}(Trolling, aggression, cyberbullying) workshop \citep{Kumar18} co-located with the International Conference of Computational Linguistics (COLING 2018) redefine hate speech detection task in terms of three type of aggression namely: Non-Aggression (NAG), Overly-Aggression(OAG) and Covertly Aggression (CAG).

\subsection{Sentiment Analysis}
During the initial year, there is a lack of standard dataset for comparative performance analysis. International Workshop on Semantic Evaluation 2013 (SemEval-2013) \citep{hltcoe2013semeval} was the first forum who developed standard tweet dataset for the benchmarking of the various sentiment analysis system. Most of the team who had participated in the competition used supervised approaches based on SVM, Naive Bayes, and Maximum Entropy. some of the team had used ensemble classifier and rule-based classifier. \citet{mohammad2013nrc} was the top team of the Semeval-2013 challenge. They have incorporated various semantic and lexicon based sentiment features for the experiment and SVM was used for the classification. Deep learning and word embedding had shown its footprints in  SemEval-2015 \citep{rosenthal2015semeval}. Team UNITN \citep{severyn2015unitn} was the second team in the message polarity task. They have build convolution neural network for the sentiment classification. They have used an unsupervised neural language model to initialize word embeddings that are further tuned by deep learning model on a distant supervised corpus \citep{severyn2015unitn}. In fourth edition SemEval-2016 \citep{nakov2016semeval},Team SwissCheese \citep{deriu2016swisscheese} was the first ranked team with $F_1$ score around 63.3 \%. Their approach was based on 2-layer convolution neural networks whose predictions are combined using a random forest classifier. SemEval-2017 \citep{rosenthal2017semeval} was the fifth edition, Team DataStories \citep{baziotis2017datastories} was the top-ranked team with AvgRec= 68.1 and $F_1$ around=67.7 \%. They use Long Short-Term Memory (LSTM) networks augmented with two kinds of attention mechanisms, on top of word embedding pre-trained on a big collection of Twitter messages without using any hand-crafted features.

\subsection{Hate Speech/Cyberbullying/Aggression Detection}
Hate Speech Detection research attracts researchers from the diverse background like Computational linguistic, computer science, social science. The actual term hate speech was coined by \citet{warner2012detecting}. Various Authors used different notion like offensive language \citep{razavi2010offensive}, Cyberbullying \citep{xu2012learning}, Aggression \citep{trac2018dataset}. \citet{davidson2017automated} studied tweet classification of hate speech and offensive language and defined hate speech as following: language that is used to expresses hatred towards a targeted group or is intended to be derogatory, to humiliate, or to insult the members of the group. Authors observed that offensive language often miss-classified as hate speech. They have trained a multi-class classifier on N-gram features weighted by its TF-IDF weights and PoS tags. In addition to these, features like sentiment score of each tweet, no of hashtags, URLS, mentions are considered. Authors concluded that Logistic regression and Linear SVM  perform better than NB, Decision Tree, Random Forests. \citet{schmidt2017survey} perform comprehensive survey on hate speech. They have identified features like Surface features, sentiment, word generalization,lexical, linguistics etc. can be used by classifier.

Cyberbullying is the type bullying that occurs on social media platform or app via cellphone or any internet enabled device. \citet{xu2012learning} introduces Cyberbullying to the NLP community. They have performed various binary classification on tweets text with bullying perspective to determine whether the user is cyberbully or not. They reported binary classification accuracy around 81\%. \citet{kwok2013locate}, authors performed classification using NB classifier  on tweets based on two classes :racist and non-racists and achieved accuracy around 76 \%.  \citet{burnap2015cyber}, authors studied cyber hate on Twitter. They have used various classifier like SVM, BLR, RFDT, Voting base ensemble for the binary classification achieved best F1-score of 0.77 in the voted ensemble.\citet{malmasi2017detecting}, authors have used NLP based lexical approach to address the multi-class classification problem. They have used character N-gram, word N-gram and word skip-gram feature for the classification. 

\citet{schmidt2017survey}, have described the key areas that have been explored to detect hate speech. They have surveyed different types of features used for hate speech classification. They have categorized features in Simple surface features, word generalization features, sentiment features, linguistic features, lexical resources features, Knowledge-based features, and Meta-Information features Simple surface features include features like character level unigram/n-gram, word generalization features include features like the bag-of-words, clustering, word embedding, paragraph embedding. Linguistic features include PoS tag of tokens. list of bad words or hate words can be considered as a lexical resource. \citet{malmasi2018challenges}, tried to address the problem of discriminating profanity from the hate speech in the social media posts. n-grams, skip-gram and clustering based word representation features are considered for the 3-class classification.The Author use SVM and advance ensemble based classifier for this task and achieved 80 \% accuracy.

\citet{aroyehun2018aggression} performed  translation as data augmentation strategy. TRAC Dataset \citep{trac2018dataset} was also augmented using translation and pseudo labeled using an external dataset on hate speech. they have reported best performance with LSTM and $F1$ score around 0.6415 on TRAC English dataset \citep{trac2018dataset}. \citet{arroyo2018cyberbullying} implement ensemble of the Passive-Aggressive (PA) and SVM classifiers with character n-grams. TF-IDF weighting used for feature representation. FIRE initiative also gave importance text representation in Indian language since its inception. \citep{majumder2008text}\citep{majumdar2007initiative}.

\section{Dataset}
\label{Dataset}
Experiments are performed on standard benchmarked Datasets to evaluate the performance of various text representation scheme. For User Aggression detection problem, Trolling, Aggression and Cyberbullying TRAC \citep{trac2018dataset} is considered for the experiments which contain post in English and code-mixed Hindi. For the Factual Detection task, experiments are performed on FIRE  IRMiDis Dataset.
 
\subsection{TRAC Dataset}
TRAC (Trolling, Aggresion and Cyberbullying) consist of 15,001 aggression-annotated Facebook Posts and Comments each in Hindi (Romanized and Devanagari script) and English for training and validation \cite{trac2018dataset}. 
%Table \ref{tab:class-stats}  shows a detail description of the training and validation Dataset. Table \ref{tab:test-data-stats} gives details of the test data corpus which contain surprise Dataset from the Twitter.

\begin{table}
\tbl{Class distribution in the Training Dataset}
{\begin{tabular}{lcccc} \toprule
 & \multicolumn{2}{c}{English Corpus} & \multicolumn{2}{c}{Hindi Corpus} \\ \cmidrule{2-5}
 &\bf \# Training  & \bf \# Validation & \bf \# Training  & \bf \# Validation  \\  \midrule  \\
 NAG & $5{,}052$ & $1,233$ & $2,275$ & $538$  \\
 CAG & $4240$ & $1,057$ & $4,869$ & $1,246$\\
 OAG & $2,708$ & $711$ & $4,856$ & $1217$\\
 Total & $12,000$ & $3,001$ & $12,000$ & $3,001$ \\ \bottomrule
\end{tabular}}
%\tabnote{\textsuperscript{a}This footnote shows how to include
 %footnotes to a table if required.}
\label{tab:class-stats}
\end{table}

\begin{table}
\tbl{Test Data Corpus statistics}
{\begin{tabular}{ll} \toprule

 \bf Test Dataset &\bf \# of posts  \\ \midrule

Facebook English Corpus & $916$ \\

Twitter English Corpus & $1,257$ \\

Facebook mixed script Hindi Corpus & $970$ \\

Twitter mixed script Hindi Corpus & $1,194$ \\ \bottomrule

\end{tabular}}
%TODO : you should write a descriptive caption
%\caption{ Test Data Corpus statistics}
\label{tab:test-data-stats}
\end{table}

\subsection{FIRE IRMiDis Dataset}
Forum for Information Retrieval Evaluation,  have introduced Microblog track since 2016 as Information Retrieval from Microblogs during Disasters (IRMiDis). IRMiDis track \citep{fire2018-irdimis}  of   FIRE %\footnote{http://fire.irsi.res.in/fire/2018/home} 
is organized with the objective to extract factual or fact-checkable tweets during the disaster which might be helpful to the victims or the people who are involved in the relief operation. Dataset contain tweets which are downloaded from the Twitter during Nepal earthquake 2015.  Following are the example of factual or fact-checkable and non-fact-checkable tweet.Table \ref{tab:class-stats-Fire} shows a detail statistics of FIRE IRMiDis Dataset. As we look at the table, There are only 83 tweets is annotated with objective class. not a single tweet is annotated from the subjective class.

\begin{table}
\tbl{FIRE IRMiDis Dataset statistics}
{\begin{tabular}{lll} \toprule
 
 \bf Particulars & \# tweets &Remark \\ \midrule

%Random Baseline & 0.3206 \\
%\hline
Number of Tweets & $50000+$ \\

Labelled Tweets & $83$  & only tweets belong to Factual class\\

Classes & 2 \\ \bottomrule

\end{tabular}}
%TODO : you should write a descriptive caption
%\caption{FIRE IRMiDis Dataset statistics}
\label{tab:class-stats-Fire}
\end{table}

\section{Text Representation Schemes}
\label{Text-rep-scheme}
The main objective of this paper is an identification of the best text representation scheme for the Social media text which is very sparse and noisy in nature. Text representation is about representing documents in a numerical way so that they can be feed as an input to the classifier. This numerical representation is in the form of the vectors which together form matrices. Essentially, There are two types of text representation scheme :(i) Bag-of-words(BoW) (ii) Distributed Word/sentence representation. BoW with count vector and TF/IDF weighting , various word embedding techniques(Word2Vec, Glove, fastText), and sentence or paragraph embedding (Doc2Vec) are studied.

\subsection{Bag-of-Word Model for Text Representation}
The Bag-of-words is the simple technique to represent the document or social media posts in the vector form and also a very common feature extraction method from the text. Word count or TF/IDF weight of each n-gram word can be used as a features. The dimension of the vector is equal to the size of vocabulary of the text corpus or dataset which results in very high dimensional sparse document vector. It is the common method used for the text representation in order to perform various NLP task like text classification, clustering. However,the BoW methods ignore the word order which may lead to loss of the context.

\subsection{Word Embedding for Text representation}
Word Embedding is the text representation technique to represent the word in the low dimensional space so that semantically similar word have similar representation. Major word embedding techniques like Word2vec learn word embedding using shallow neural network. The fastText, extension of Word2vec, consider the morphological structure of the word. 

\subsubsection{Word2Vec}
Word2vec \citep{mikolov2013efficient} is the unsupervised and predictive neural word embedding technique to learn the word representation in the low dimensional space. Word2vec is a two-layer neural net that take text corpus as an input and output is a set of vectors. two novel model architectures: Skipgram and CBOW(Continuous bag of words)  are proposed for computing continuous vector representations of words from very large data sets.

\subsubsection{Glove}
GloVe stands for Global vector for [Word Representation] \citep{pennington2014glove}is an unsupervised method for learning word embedding. A Co-occurrence word matrix is created from the text corpus for the training and is reduced in low dimensional space which explain the variance of high dimensional data and provide word vector for each word.

\subsubsection{fastText}
fastText \citep{bojanowski2017enriching} is the neural word embedding technique which learn distributed low dimensional word embedding. Word2vec, Glove consider each word as single unit and ignore the morphological structure of the word. They are not able generate word embedding for the unseen or out of vocabulary word during the training. fastText overcome this limitation of Word2vec and GLOVE by considering each word as N-gram of characters. A word vector for a word is computed from the sum of the n-gram characters. The range of N is typically 3 to 6. Since user on social media often make spelling error, typos, fastText will be more effective  then rest of two. 

\subsection{Paragraph vector/Doc2vec}
Paragraph Vector is an unsupervised algorithm that learns fixed-length feature representations from variable-length pieces of texts, such as
sentences, paragraphs, and documents \citep{le2014distributed}. Paragraph vector represents each document by a dense vector which is trained to predict words in the document. Authors believe that Paragraph vector have the potential to overcome the weaknesses of bag-of-words models and claimed that  Paragraph Vectors outperform bag-of-words models as well as other techniques for text representations. Paragraph vector model is also referred as doc2vec model. Henceforth, we will refer paragraph and Doc2vec interchangeably. Doc2vec model have two architecture namely : (i) DM: This is the Doc2Vec model analogous to CBOW model in Word2vec. The paragraph vectors are obtained by training a neural network on the task of inferring a center word based on context words and a context paragraph. (ii)  DBOW: This is the Doc2Vec model analogous to Skip-gram model in Word2Vec. The paragraph vectors are obtained by training a neural network on the task of predicting a probability distribution of words in a paragraph given a randomly-sampled word from the paragraph.

\subsection{Transfer Learning }
Transfer Learning in NLP is not as matured as compare to in Computer Vision. Transfer learning is a method in which model is trained on large corpus for a particular task and use this pre-trained model for the similar task. There are two way to use transfer learning in NLP (i) Use of Pre-trained word embedding to initialize first layer of neural network model which can be called as shallow representation. (ii) Use the full model and fine tune for the task specific in supervise learning way.

%\subsubsection{Pre-Trained Word vector}
Word2vec, Glove and fastText provide pre-trained word vector trained on the large corpus. Google Word2vec pre-trained model have word vector for 3 million words with size 300  and trained on Google news. Glove pre-trained model available with different embed size and trained on common crawl, Twitter. We have use Glove pre-trained model with vocabulary size 2.2 million and trained on common crawl. fastText pre-trained models are available in 157 language. We have use fasttext pre-trained vector for Englsih and Hindi language trained on commnon crawl and wikipedia.

%\subsubsection{Pre-trained Language Model}
Recently, transfer learning in NLP done in new way; First language model is trained on large text corpus in unsupervise way and fine tune on specific task like text classification on labeled data. \citet{peters2018deep} author argued that word representation is depend upon the context. So each word has different word vector depending upon the position of the word in the sentence. Essentially Each word has dynamic word vector with respect to the context as opposed to the traditional word embedding techniques which always give same word vector ignoring the context. Embedding from Language Models (ELMos) use languge model for the word embedding.  \citet{howard2018universal} author propose Universal Language Model Fine-Tuning for Text Classification (ULMFiT) which is bi-LSTM model that is trained on a general language modeling (LM) task and then fine tuned on text classification. Results are reported on both transfer learning model on TRAC dataset \citep{trac2018dataset}.

\section{Evaluation Tasks}
\label{task}
We have benchmarked  various text representation scheme on  two specialized NLP task namely: aggression detection and fact detection. Text Representation scheme are evaluated on machine learning and deep neural model.

\subsection{Aggression Detection task}
The objective of this task is to identify type of aggression present in the text in both Englsih and code-mixed Hindi language. Aggression are classified into three level namely: `Overtly Aggressive` (OAG), `Covertly Aggressive` (CAG) and `Non-aggressive` (NAG). We have implemented all standard machine learning classifiers like Multinomial Naive Bayes (MNB), Logistic Regression(LR), K-Nearest neighbors (KNN), Support Vector Classifier (SVC), Decision Tree (DT),Stochastic Gradient Descent(SGD), Random forest (RF),  Ridge, AdaBoost, Perceptron, and various voting based ensemble with different text representation schemes like count based, TF/IDF and word embedding to prepare baseline results. Various word embedding techniques like Word2Vec, Glove, fastText, Paragraph2Vec are studied. 

\subsubsection{Problem statement}
Basically Aggression detection is a Text classification problem. Formally, the task of Text Classification is stated as follows. Given a set of social media feed  and a set of classes, We need to compute a function  of the form: \[ C=f(T,\Omega) \]

where f is the multi-class classifier that is computed using training data, T is the numeric representation of the text of the dataset, $\Omega$ is the set of parameters of the classifier and C is the pre-define class-labels.

\subsubsection {Model Architectures and Hyperparameters}
In this subsection, we will discuss the architecture and hyperparameters of our deep neural model used for the classification. Model learns feature from the input texts Ther is no need to design hand-crafted features which used to encode text into feature vector.  

\paragraph{Bidirectional LSTM}
The first model is based on the Bidirectional LSTM  include embedding layer with embed size 300, convert each word from the post into a fixed length vector. short posts are padded with zero values. Subsequent layers includes Bidirectional LSTM layer with 50 memory units followed by one-dimensional global max pooling layer, a hidden layer with size 50  and output layer with softmax activations. ReLU activation function is used for the hidden layer activation. A drop out layer is added between the last two layers to counter the overfitting with parameter 0.1. Hyperparameters are as follows: Sequence length is fixed at 1073 word; maximum length of posts in the dataset. No of features is equal to half of total vocabulary size. Models are trained for 10 epoch with batch size 128. Adam optimization algorithm is used to update network weights.

\paragraph{Single LSTM with higher dropout}
This model is based on the Long Short Term Memory, a type of recurrent neural network with higher dropout. This model is having one embedding layer, one LSMT layer with a size 64 memory unit, and one fully connected hidden layer with Relu activation and size 256  and an output layer with softmax activation.  Hyperparameters are same as discussed in the previous model. A dropout layer is added between the hidden layer and an output layer with drop out rate 0.2  to address the overfitting issue.

\paragraph{CNN Model}
This model includes one embedding layer whose weights are initialized with fastText pre-trained vector with embed size is 300, followed by one-dimensional convolution layer with 100 filters of height 2 and stride 1 to target biagrams. In addition to this, Global Max Pooling layer added to fetch the maximum value from the filters which are feed to the fully connected hidden layer with size 256, followed by output layer. ReLU and softmax activation function are used for the hidden layer and output layer respectively.

\paragraph{CNN model with Multiple Convolution layer}
This model includes embedding layer with embed size 300. Three one dimensional convolution layers with size 100 and different filters with height 2,3,4 to target bigrams, trigrams, and four-grams features, followed by max pooling layer which concatenate max pooled result from each of one-dimensional convolution layer. The final two layers include a fully connected hidden layer with size 250 and output layer with ReLu and softmax activation. A Drop out layer is added between the last two layer with rate 0.2. Hyperparameters are same as discussed in the first model. This model is similar to proposed by \citep{zhang2015sensitivity}.

\subsection{Factual Post/Tweet Detection from Social Media}
During the emergency situation like earthquake or floods, Microblog plays a very important role as an anonymous communication medium. The various entity like, Volunteers, NGOs involved in relief operation always look for real-time information which contains facts instead of prayer and condolence messages. In more technical term, these agencies are looking for factual information from Microblog instead of the subjective information. In addition to this, the system should generate  rank-list of the tweets  based upon the worthiness of facts. we considered this problem as a binary classification problem plus pure IR Ranking problem. two classes can be labeled as factual and non-factual. 

the IRMiDis dataset  \footnote{https://sites.google.com/site/irmidisfire2018/},which was prepared from the tweet posted during Nepal earthquake 2015 \cite{fire2018-irdimis} is considered for the experiment. There are only 83 fact checkable tweets in the dataset. Non-factual tweets are not available. Total no of tweets in the dataset is more than 50000. 

\subsubsection{Preparation of Training Data }
Due to the unavailability of adequate training data, The first task is to prepare training data to train the deep neural model. We randomly choose 100 tweets from the dataset and labeled as a  non-fact-checkable tweet and 83 fact-checkable tweets present in the dataset labeled as fact-checkable. We have trained our Convolution neural network on these training data and tested the model on the remaining 50000 tweets. At this stage we are not interested in the class but, we have sorted all the tweets based upon the predicted probability of the fact-checkable class and selected top 2000 tweets. We have randomly selected tweets and gave relevance judgment based upon availability of factual information in first 1000 tweets and manually extracted 300 tweets as non-fact-checkable tweets to minimize the false positives. Remaining 1700 tweets labeled as fact-checkable tweets. We selected the last 1700 tweets with the least probability of the class fact-checkable and labeled them as non-fact-checkable tweets. So our Training corpus has 1783 fact-checkable and 2000 non-fact-checkable tweets.
\\
\subsubsection{Proposed Approach}
We have used word embedding to represent the text instead of bags-of-words. fastText \citep{mikolov2018advances} pre-trained vector with 300 dimensions is used to initialize the weight matrix of the embedding layer of the network. We trained our CNN model on this training corpus with 10-fold cross-validation.The Model gives validation accuracy around 94\%. Finally, we run the model on the entire corpus and sorted the tweet based upon the predicted probability of the Fact-checkable class. Essentially this approach termed as weakly-supervise classification.

\section{Results}
\label{sec:results}

In this section, we first present results of classifiers TRAC dataset \cite{trac2018dataset} with different text representation scheme. Latter we present result on FIRE IRMiDis 2018 Dataset. Tweets are very noisy in nature contains user mentions, Hashtags, Emojis, and URLs. We do not perform any kind of text pre-processing on tweets in experiments with deep neural models. In experiments with machine learning classifier, before classification, Hashtag symbol \# and User mentions are dropped from the tweets. Non-ASCII characters and stop-words are removed from tweet text \citep{modha2016daiict}.

\subsection{Results On TRAC Dataset}

\begin{figure}
\centering

\includegraphics[scale=0.45]{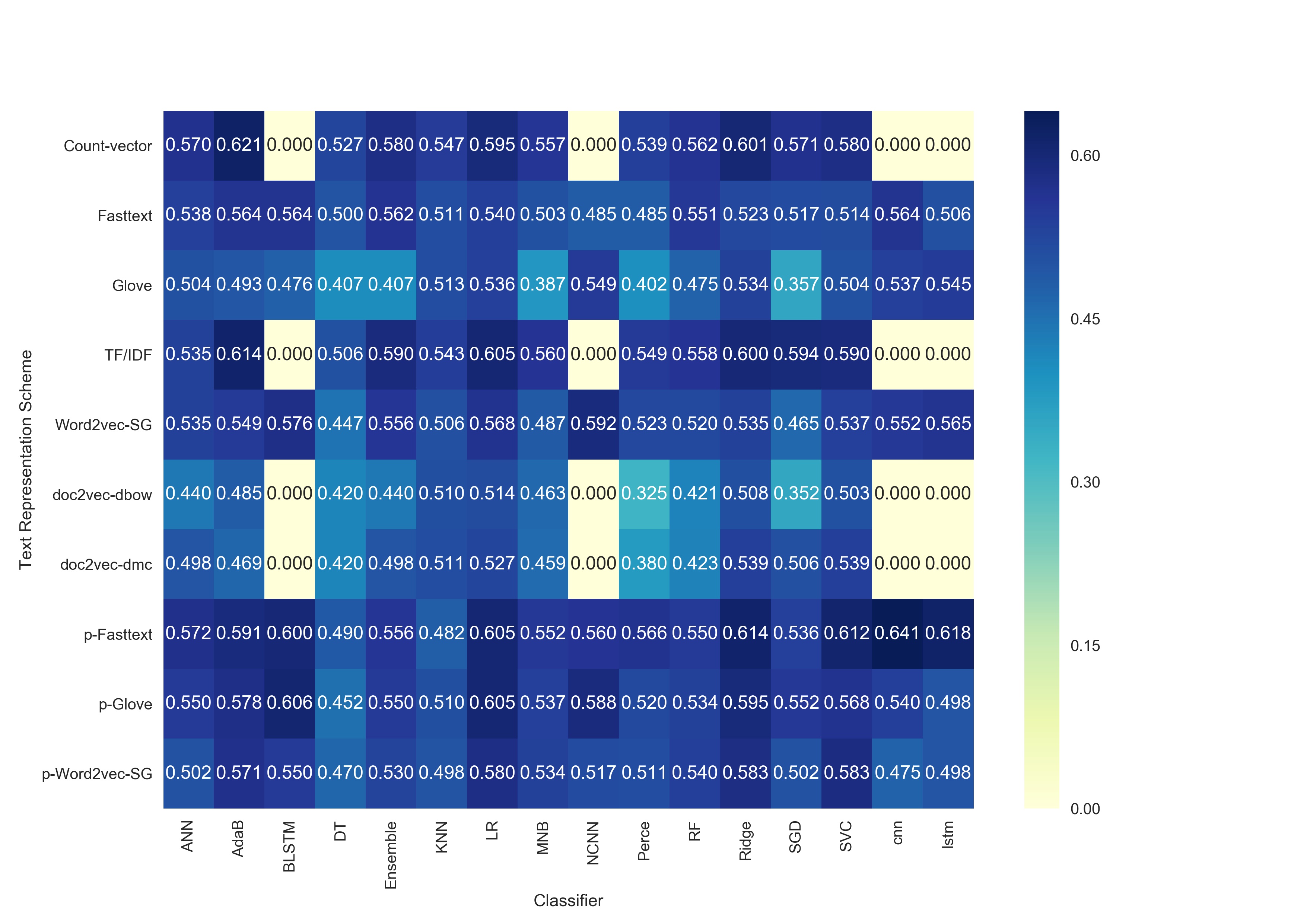}
%\subfloat[A slightly shorter sub-caption.]{%
%\resizebox*{5cm}{!}{\includegraphics{graph2.eps}}}
\caption{Heatmap on English Facebook Test Dataset Results.} \label{fig:hf}
\end{figure}

\begin{figure}
\centering

\includegraphics[scale=0.45]{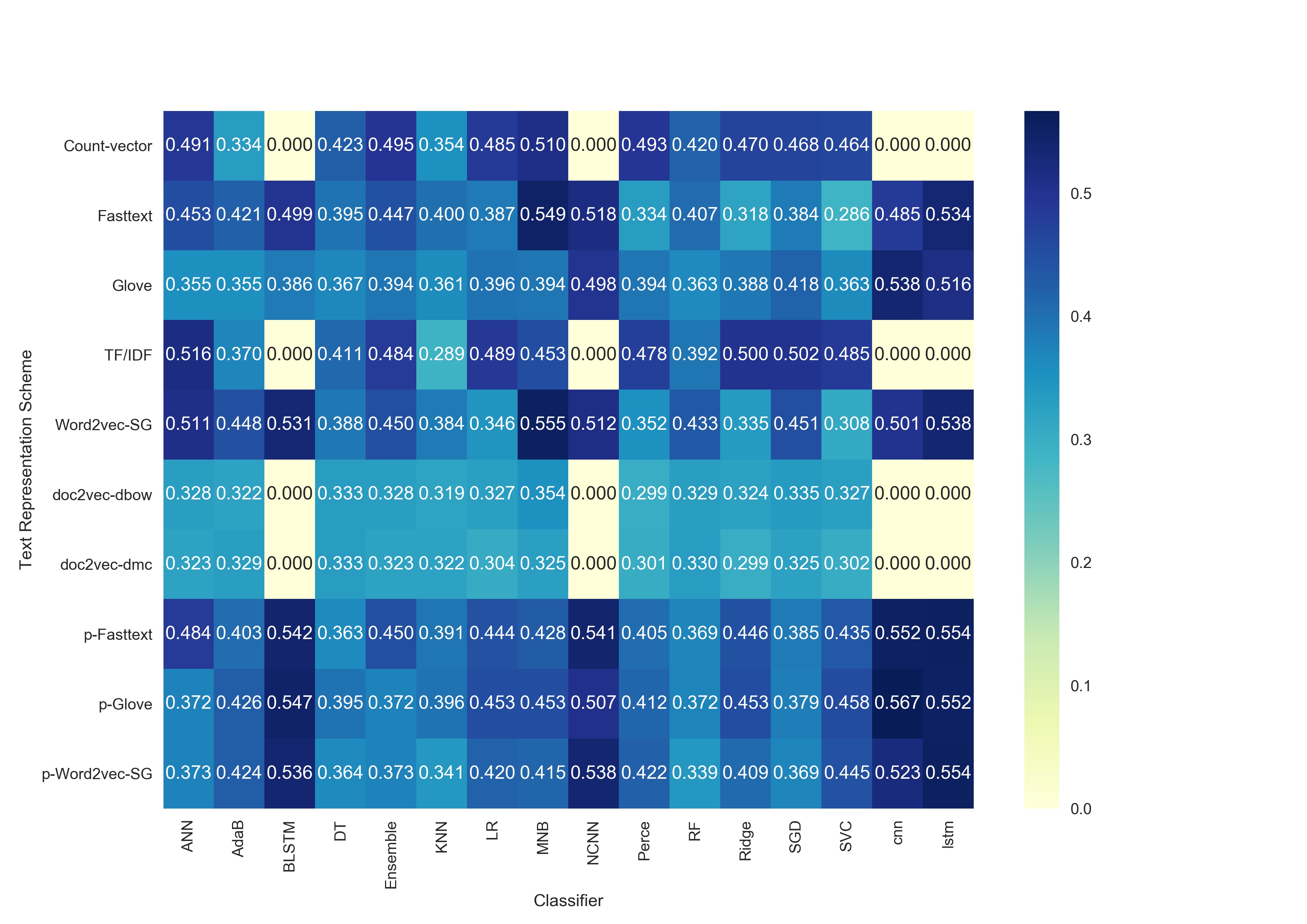}
%\subfloat[A slightly shorter sub-caption.]{%
%\resizebox*{5cm}{!}{\includegraphics{graph2.eps}}}
\caption{Heatmap on English Twitter Test Dataset Results.} \label{fig:ht}
\end{figure}

Precision, Recall, and F1-score are the standard metrics which are used to evaluate the classifier performance. We have evaluated 16 classifiers performance on 4 Datasets (2 English+2 Hindi) with 10 Text Representation scheme(8 in the case of Hindi Dataset). Looking at such massive experiment, it is difficult to report  results in all the above metrics. Therefore, Results are reported in terms of weighted F1-score only which is the function of Precision and Recall. Classifiers results based on LSTM and CNN on BoW text representation schemes are not possible due to the high dimensionality.  Bernoulli classifier is used instead of  Naive Bayes Classifier in case of text representation schemes other than BoW. Since word vectors might have negative weights, it is impossible to calculate probabilities with negative weights. Skip-gram variant of Word2Vec and fastText is used in this experiment instead of continuous bag-of-word.
Table \ref{tab:res-en-fb-mc} and \ref{tab:res-en-tw-mc} shows results on Facebook and Twitter English Dataset with BoW and word embedding while Table \ref{tab:res-en-fbtw-dnn} present result with pre-trained word embedding with same dataset. Table \ref{tab:res-hi-fb-mc} and \ref{tab:res-hi-tw-mc} shows results on Facebook  and Twitter code-mixed Hindi Dataset. Only fastText provide pre-trained word vector \citep{mikolov2018advances} for the Hindi language. Exhaustive evaluation is performed with all classifiers with respect to each text representation schemes. Experiments are also performed with the new transfer learning model like ELMO and ULMFIT. Table \ref{tab:res:transfer} presents results on both Facebook and Twitter English Datasets. Figure \ref{fig:hf} and figure \ref{fig:ht} display the heatmap of the results achieved by classifiers on each text representation scheme.

\begin{table}
\tbl{F1-score on TRAC Facebook English Dataset}
% Table generated by Excel2LaTeX from sheet 'Hindi-summ'
%\begin{tabular}{|l|l|l|l|l|l|l|l|l|}
{\begin{tabular}{p{1.4 cm}p{1cm}p{1 cm}p{1 cm}p{1 cm}p{1.3 cm}p{1.5 cm}p{1.2 cm}} \toprule

{\bf Classifier} & {\bf Count-vector} & {\bf TF/IDF} & {\bf W2Vec} & {\bf Glove} & {\bf Fasttext}  & {\bf doc2vec-dmc} & {\bf doc2vec-dbow} \\ \midrule

 {NB} &     0.5571 & {\bf 0.5596} &     0.4870 &     0.3873 &     0.5035 &     0.4585 &     0.4634 \\ 
  {LR} &     0.5953 & {\bf 0.6046} &     0.5675 &     0.5358 &     0.5400 &     0.5266 &     0.5139 \\

 {KNN} &     0.5466 & {\bf 0.5428} &     0.5061 &     0.5130 &     0.5113 &     0.5114 &     0.5095 \\

 {SVC} &     0.5801 & {\bf 0.5902} &     0.5369 &     0.5037 &     0.5137 &     0.5388 &     0.5033 \\

  {DT} & {\bf 0.5269} &     0.5055 &     0.4468 &     0.4067 &     0.5002 &     0.4198 &     0.4198 \\

 {SGD} &     0.5706 & {\bf 0.5938} &     0.4647 &     0.3571 &     0.5167 &     0.5060 &     0.3521 \\
  { RF} & {\bf 0.5621} &     0.5582 &     0.5199 &     0.4752 &     0.5513 &     0.4230 &     0.4210 \\
{Ridge} & {\bf 0.6009} &     0.5999 &     0.5347 &     0.5336 &     0.5225 &     0.5385 &     0.5083 \\
{AdaB} & {\bf 0.6210} &     0.6141 &     0.5491 &     0.4932 &     0.5644 &     0.4689 &     0.4852 \\
{Perce.} &     0.5387 & {\bf 0.5491} &     0.5230 &     0.4020 &     0.4848 &     0.3800 &     0.3253 \\
 {ANN} & {\bf 0.5703} &     0.5350 &     0.5350 &     0.5037 &     0.5380 &     0.4980 &     0.4401 \\
{Ensemble} &       0.58 & {\bf 0.5900} &     0.5558 &     0.4067 &     0.5617 &     0.4980 &     0.4401 \\

  LSTM & & &  \bf   0.5649 &     0.5454 &     0.5062 &  &    \\
  BLSTM &  & &   \bf  0.5759 &     0.4760 &     0.5641 &  &    \\
  CNN & &   &     0.5515 &     0.5365 & \bf     0.5638 &  &   \\
  NCNN &  &   &   \bf  0.5919 &     0.5488 &     0.4849 &  &   \\ \bottomrule

\end{tabular}}  

%\caption{F1-score on TRAC Facebook Code-mixed Hindi Dataset}
\label{tab:res-en-fb-mc}
\end{table}

\begin{table}
\tbl{F1-score on TRAC Twitter English Dataset}
{\begin{tabular}{p{1.4 cm}p{1cm}p{1 cm}p{1 cm}p{1 cm}p{1.3 cm}p{1.5 cm}p{1.2 cm}} \toprule

{\bf Classifier} & {\bf Count-vector} & {\bf TF/IDF} & {\bf W2Vec} & {\bf Glove} & {\bf Fasttext}  & {\bf doc2vec-dmc} & {\bf doc2vec-dbow} \\ \midrule

NB &     0.5102 &     0.4528 & {\bf 0.5551} &     0.3936 &     0.5495 &     0.3254 &     0.3536 \\

        LR &     0.4849 & {\bf 0.4890} &     0.3457 &     0.3959 &     0.3871 &     0.3041 &     0.3274 \\

       KNN &     0.3539 &     0.2891 &     0.3843 &     0.3607 & {\bf 0.3997} &     0.3225 &     0.3191 \\

       SVC &     0.4642 & {\bf 0.4853} &     0.3078 &     0.3627 &     0.2858 &     0.3019 &     0.3274 \\

        DT & {\bf 0.4229} &     0.4111 &     0.3884 &     0.3673 &     0.3948 &     0.3326 &     0.3326 \\

       SGD &     0.4682 & {\bf 0.5020} &     0.4512 &     0.4182 &     0.3838 &     0.3251 &     0.3350 \\
       
        RF &     0.4199 &     0.3917 & {\bf 0.4333} &     0.3634 &     0.4069 &     0.3301 &     0.3293 \\

     Ridge &     0.4703 & {\bf 0.5003} &     0.3352 &     0.3877 &     0.3180 &     0.2994 &     0.3243 \\

      AdaB &     0.3343 &     0.3696 & {\bf 0.4485} &     0.3552 &     0.4215 &     0.3288 &     0.3223 \\

    Perce. & {\bf 0.4930} &     0.4778 &     0.3521 &     0.3938 &     0.3340 &     0.3015 &     0.2990 \\

ANN & {\bf 0.4912} & {\bf 0.5164} &     0.5111 &     0.3552 &     0.4532 &     0.3230 &     0.3281 \\

  Ensemble & {\bf 0.495} &     0.4842 &     0.4500 &     0.3938 &     0.4471 &     0.3230 &     0.3281 \\

      LSTM &            &            &   \bf  0.5385 &     0.5156 &     0.5335 &            &            \\

     BLSTM &            &            &   \bf  0.5314 &     0.3860 &     0.4985 &            &            \\
       CNN &            &            &     0.5012 & \bf    0.5377 &     0.4849 &            &            \\

      NCNN &            &            &     0.5120 &     0.4984 & \bf     0.5179 &            &            \\ \bottomrule
\end{tabular}}  

%\caption{F1-score on TRAC Facebook Code-mixed Hindi Dataset}
\label{tab:res-en-tw-mc}
\end{table}

\begin{table}

\tbl{F1-score on TRAC Facebook and Twitter English Dataset:Using Pre-trained word vectors.}
% Table generated by Excel2LaTeX from sheet 'Hindi-summ'
{\begin{tabular}{lcccccc}\toprule

           & \multicolumn{ 3}{c}{{\bf Facebook Test Dataset}} & \multicolumn{ 3}{c}{{\bf Twitter Test Dataset}} \\ \cmidrule{2-6}

{\bf Classifier} & {\bf p-Word2vec} & {\bf p-Glove} & {\bf p-Fasttext} & {\bf p-Word2vec} & {\bf p-Glove} & {\bf p-Fasttext} \\ \midrule

  {NB} &     0.5342 &     0.5373 & {\bf 0.5519} &     0.4152 & {\bf 0.4527} &     0.4276 \\

  {LR} &     0.5799 &     0.6050 & {\bf 0.6045} &     0.4197 & {\bf 0.4527} &     0.4441 \\
 {KNN} &     0.4981 & {\bf 0.5103} &     0.4819 &     0.3405 & {\bf 0.3959} &     0.3912 \\
 {SVC} &     0.5832 &     0.5678 & {\bf 0.6120} &     0.4446 & {\bf 0.4581} &     0.4350 \\
  {DT} &     0.4700 &     0.4515 & {\bf 0.4900} &     0.3640 & {\bf 0.3949} &     0.3632 \\
 {SGD} &     0.5019 &     0.5521 & {\bf 0.5360} &     0.3692 &     0.3793 & {\bf 0.3852} \\

  {RF} &     0.5402 &     0.5338 & {\bf 0.5505} &     0.3394 & {\bf 0.3716} &     0.3687 \\

{Ridge} &     0.5829 &     0.5952 & {\bf 0.6140} &     0.4092 & {\bf 0.4530} &     0.4461 \\

{AdaB} &     0.5713 &     0.5781 & {\bf 0.5907} &     0.4241 & {\bf 0.4261} &     0.4033 \\
{Perce.} &     0.5114 &     0.5201 & {\bf 0.5660} & {\bf 0.4224} &     0.4118 &     0.4049 \\
 {ANN} &     0.5025 &     0.5498 & {\bf 0.5722} &     0.3728 &     0.3722 & {\bf 0.4842} \\
{ Ensemble} &     0.5300 &     0.5500 & {\bf 0.5558} &     0.3728 &     0.3722 & {\bf 0.4500} \\
{LSTM} &     0.4979 &     0.4979 & {\bf 0.6178} &     0.5537 &     0.5518 & { 0.5541} \\

{ BLSTM} &     0.5501 & {\bf 0.6062} &     0.6000 &     0.5359 & {\bf 0.5466} &     0.5423 \\
 { CNN} &     0.4749 &     0.5405 & {\bf 0.6407} &     0.5226 & {\bf 0.5667} &     0.5520 \\

{ NCNN} &     0.5169 & {\bf 0.5883} &     0.5600 &     0.5384 &     0.5067 & {\bf 0.5407} \\ \bottomrule

\end{tabular}}

%\caption{Pre-trained model:F1-score on TRAC Facebook and Twitter English Dataset}
\label{tab:res-en-fbtw-dnn}
\end{table}

\begin{table}
\tbl{F1-score on TRAC Facebook Code-mixed Hindi Dataset}
% Table generated by Excel2LaTeX from sheet 'Hindi-summ'
%\begin{tabular}{|l|l|l|l|l|l|l|l|l|}
{\begin{tabular}{p{1.4 cm}p{1cm}p{1 cm}p{1 cm}p{1 cm}p{1.3 cm}p{1.5 cm}p{1.2 cm}p{1.2 cm}} \toprule

{\bf Classifier} & {\bf Count-vector} & {\bf TF/IDF} & {\bf W2Vec} & {\bf Glove} & {\bf Fasttext} & {\bf p-fastText} & {\bf doc2vec-dmc} & {\bf doc2vec-dbow} \\ \midrule

  { NB} &     0.5535 & {\bf 0.6031} &     0.3001 &      0.372 &     0.2959 &     0.3176 &     0.3459 &     0.4736 \\

  { LR} &     0.5855 & {\bf 0.6134} &     0.5779 &      0.464 &     0.5457 &     0.5518 &     0.3894 &     0.4380 \\

 { KNN} &     0.3340 &     0.1721 &     0.4998 &      0.425 & {\bf 0.5106} &     0.4909 &     0.3768 &     0.4038 \\
 { SVC} &     0.5556 & {\bf 0.5862} &     0.4806 &      0.373 &     0.5186 &     0.5442 &     0.3879 &     0.4344 \\

  { DT} & {\bf 0.5307} &     0.5025 &     0.4629 &      0.388 &     0.4392 &     0.4288 &     0.3485 &     0.3485 \\
 { SGD} &     0.5533 & {\bf 0.5922} &     0.3912 &      0.393 &     0.3670 &     0.4746 &     0.3331 &     0.4134 \\

  {RF} &     0.5473 & {\bf 0.5473} &     0.5374 &      0.440 &     0.5047 &     0.4788 &     0.3512 &     0.3477 \\

{Ridge} &     0.5780 & {\bf 0.5850} &     0.5293 &      0.381 &     0.5092 &     0.5544 &     0.3866 &     0.4292 \\

{AdaB} & {\bf 0.5373} &     0.5233 &     0.5342 &      0.479 &     0.5336 &     0.4913 &     0.3751 &     0.4214 \\
{Perce.} &     0.5213 & {\bf 0.5598} &     0.4232 &      0.364 &     0.3763 &     0.4873 &     0.2661 &     0.3282 \\

 {ANN} & {\bf 0.5703} &     0.5350 &     0.5455 &     0.5037 & {\bf 0.5842} &     0.5190 &     0.4091 &     0.4440 \\

{Ensemble} &     0.5700 & {\bf 0.6087} &     0.5558 &     0.4067 &      0.534 &     0.5612 &     0.4980 &     0.4401 \\

      LSTM &            &            &     0.5649 &      0.590 &  \bf   0.6021 &     0.5916 &            &            \\

     BLSTM &            &            &     0.5759 &      0.527 &     0.5770 &  \bf   0.5900 &            &            \\

       CNN &            &            &     0.5515 &      0.566 &     0.5950 &   \bf  0.6081 &            &            \\

      NCNN &            &            &     0.5919 &      0.573 &     0.5912 &  \bf   0.5965 &            &            \\ \bottomrule

\end{tabular}}  

%\caption{F1-score on TRAC Facebook Code-mixed Hindi Dataset}
\label{tab:res-hi-fb-mc}
\end{table}

\begin{table}
\tbl{F1-score on TRAC Twitter Code-mixed Hindi Dataset}
% Table generated by Excel2LaTeX from sheet 'Hindi-summ'
%\begin{tabular}{|l|l|l|l|l|l|l|l|l|}
{\begin{tabular}{p{1.4 cm}p{1cm}p{1 cm}p{1 cm}p{0.9 cm}p{1.3 cm}p{1.6 cm}p{1.2 cm}p{1.2 cm}} \toprule

{\bf Classifier} & {\bf Count-vector} & {\bf TF/IDF} & {\bf W2Vec} & {\bf Glove} & {\bf Fasttext} & {\bf p-fastText} & {\bf doc2vec-dmc} & {\bf doc2vec-dbow} \\ \midrule

  { NB} &     0.2970 &     0.2902 &     0.3215 &      0.273 & {\bf 0.3359} &     0.2897 &     0.3270 &     0.3205 \\

  {LR} & {\bf 0.3787} &     0.3724 &     0.2819 &      0.279 &     0.3184 &     0.3524 &     0.2438 &     0.2833 \\

 {KNN} &     0.2527 &     0.2553 & {\bf 0.3704} &      0.334 &     0.3381 &     0.2917 &     0.3051 &     0.3299 \\

 { SVC} &     0.3781 &   \bf  0.3886 &     0.2821 &      0.261 &     0.3087 &     0.3472 &     0.2580 &     0.2905 \\

  { DT} &     0.3685 & {\bf 0.3936} &     0.3572 &      0.326 &     0.3475 &     0.3473 &     0.2988 &     0.2988 \\

 { SGD} & {\bf 0.3996} &     0.3993 &     0.2822 &      0.287 &     0.2739 &     0.3163 &     0.2605 &     0.2588 \\

  { RF} &     0.3585 & {\bf 0.3737} &     0.3286 &      0.344 &     0.3449 &     0.3288 &     0.2981 &     0.2988 \\

{ Ridge} &     0.3616 & {\bf 0.3872} &     0.2811 &      0.242 &     0.3346 &     0.3361 &     0.2549 &     0.2875 \\

{ AdaB} &     0.1886 &     0.1903 &     0.3256 & {\bf 0.362} &     0.3261 &     0.3441 &     0.2614 &     0.2933 \\

{ Perce.} & {\bf 0.3931} &     0.3868 &     0.2802 &      0.329 &     0.2787 &     0.3835 &     0.2616 &     0.2752 \\

 { ANN} &     0.430 & {\bf 0.44} &     0.3163 &     0.3552 &     0.2399 &     0.3593 &     0.2419 &     0.3132 \\

{ Ensemble} &     0.4400 & {\bf 0.4600} &     0.4500 &     0.3938 &     0.3426 &     0.3555 &     0.3230 &     0.3281 \\

      LSTM &            &            &     0.3840 &      0.376 &     0.3667 &    \bf 0.4600 &            &            \\

     BLSTM &            &            &     0.2846 &      0.318 &     0.3005 &     \bf 0.4600 &            &            \\

       CNN &            &            &     0.3323 &      0.317 &     0.2669 &    \bf 0.4992 &            &            \\

      NCNN &            &            &     0.3338 &      0.380 &     0.3494 &     \bf 0.4600 &            &            \\ \bottomrule

\end{tabular}}  

%\caption{F1-score on TRAC Facebook Code-mixed Hindi Dataset}
\label{tab:res-hi-tw-mc}
\end{table}

\begin{table}
\tbl{F1 score on TRAC Facebook English Test Dataset using Transfer Learning methods}
{\begin{tabular}{ lcc } \toprule

Transfer learning Model &\multicolumn{2}{c}{English} \\ \cmidrule{2-3}

&Facebook Dataset & Twitter Dataset \\ \midrule

ELMO &  $0.3699$ & $0.3854$ \\

ULMFiT   &$0.4725$ & $0.4664$  \\ \bottomrule

\end{tabular}}
%TODO : you should write a descriptive caption
%\caption{F1 score on TRAC Facebook English Dataset using Transfer Learning methods}
\label{tab:res:transfer}
\end{table}

\begin{table}
\tbl{weighted F1-score TRAC Test Dataset: comparison with peers}
{\begin{tabular}{ p{3cm}p{2cm}p{2cm}p{2cm} p{2cm} }\toprule

System &\multicolumn{2}{c}{English} &\multicolumn{2}{c}{Hindi}\\ \cmidrule{2-5}

&Facebook Dataset & Twitter Dataset & Facebook Dataset & Twitter Dataset\\ \midrule

Our system result & \bf 0.6407 & 0.5541 &0.6081 &\bf 0.4992\\
DA-LD-hildesheim   &0.6178 & 0.552 & 0.6081 & \bf 0.4992 \\

saroyehun & \bf 0.6425 &0.5920 &  NA &  NA\\

EBSI-LIA-UNAM
 &0.6315 &0.5715 &NA &NA\\

TakeLab &0.5920 &0.5651&  NA &  NA\\

taraka rama  &0.6008 &0.5656 & \bf 0.6420 & 0.40 \\

vista.ue &0.5812 & \bf 0.6008 & 0.5951 & 0.4829 \\

na14 & 0.5920 &0.5663 &0.6450 & 0.4853 \\ \bottomrule

\end{tabular}}
%TODO : you should write a descriptive caption
%\caption{$F_!$ score comparision with peers}
\label{tab:res:comp-peer}
\end{table}

\subsection{Information Retrieval from Microblogs during Disasters (IRMiDis) Dataset}

As discussed in the previous section \ref{intro}, This task is classification plus Ranking task. Table \ref{tab:res:IRMiDis} shows our system results on IRMiDis dataset \citep{fire2018-irdimis} along with the rest of teams. nDCG overall is the primary metric for the evaluation. Our system substantially outperforms rest of team in the most of the metrics which justifies our claim established on TRAC dataset \citep{trac2018dataset}

\begin{table}
\tbl{Results Comparison with rest of team on FIRE 2018 IRMiDis Dataset.}
{\begin{tabular}{lllllll}\toprule

\bf System &p@100 &R@1000 &MAP@100 &MAP &nDCG @100 &\bf nDCG \\  \midrule

\bf Our System & 0.4& \bf $0.2002$ &$0.0129$ & \bf $0.1471$ &$0.4021$ & \bf $0.7492$ \\

MIDAS-semiauto &$0.9600$ &$0.1148$ &$0.0740$ &$0.1345$ &$0.6007$ &$0.6899$\\

MIDAS-1 &$0.8800$ &$0.1292$ &0.0581 &0.1329 &0.5649 &0.6835\\

FAST\_NU\_Run2 &0.7000 &0.0885 &0.0396 &0.0801 &0.5723 &0.6676\\

UEM\_DataMining\ &0.6800 &0.1427 &0.0378 &0.1178 &0.5332 &0.6396\\

iitbhu\_irlab2 &0.3900 &0.0447 &0.0144 &0.0401 &0.3272 &0.6200\\ \bottomrule
\end{tabular}}
%TODO : you should write a descriptive caption
%\caption{Results Comparison with rest of team on IRMiDis Dataset.}
\label{tab:res:IRMiDis}
\end{table}

\section{Result Analysis}
\label{lab:ra}
In this section, we will present the comprehensive result analysis and try to answer the research questions which framed before the experiments were performed. As we look at the table \ref{tab:res-en-fb-mc} \ref{tab:res-en-tw-mc}, and \ref{tab:res-en-fbtw-dnn}, Overall, LSTM and CNN with pre-trained fastText word embedding marginally outperform (around 2 \% to 4\%) standard machine learning classifiers and ensemble of classifier  with respect to weighted $F_1$- score on Facebook English corpus and substantially outperforms on Twitter English corpus. By and large similar results observed on code-mixed Hindi corpus as shown in table \ref{tab:res-hi-fb-mc} and \ref{tab:res-hi-tw-mc}.

\begin{table}
\tbl{Results Comparison with CNN model and Logistic Regression TRAC Facebook English Test Dataset}
{\begin{tabular}{llllllll}\toprule

Class &\multicolumn{3}{c}{CNN model} &\multicolumn{3}{c}{Logistic Regression} & \#Posts \\ \cmidrule{2-8}

&P &R &Weighted F1 &P &R &Weighted F1 & \\ \midrule
NAG &0.86&0.64&0.73 &0.83 &0.60 &0.70 &630 \\
CAG &0.28  &0.46 &0.35 &0.23  &0.54 &0.32   &142\\
OAG &0.42  &0.61 &0.50  &0.46   &0.39   &0.42   &144\\\bottomrule

overall &0.70  &0.61 &0.64 &0.68      &0.56      &0.60       &916 \\ \bottomrule
\end{tabular}}
\label{tab:res-ana-trac-fb-en}
\end{table}

Table \ref{tab:res-ana-trac-fb-en} present the detailed comparative results of two classifiers: CNN model with fastText pre-trained vector and the logistic regression with TF/IDF weighting on TRAC Facebook English dataset. The CNN Model classify Facebook posts better than logistic regression at the individual class level and overall. It has been quite evident that posts belong to CAG class are hard to classify and \citet{malmasi2017detecting} reported that the same observation. Table \ref{tab:res-ana-cnn-cag} show posts which are miss-classified by logistic regression however, CNN model correctly classified them into the CAG class.

\begin{table}
\tbl{sample post for the CAG class.}
{\begin{tabular}{p{0.6cm} p{7cm}p{1.6cm} p{0.6cm} p{0.6cm}} \toprule
 
no   &Post text & Gold Label & CNN  & LR \\ \midrule
1 & Mauni singh trying very hard to convince himself what is written in script... body language says it all & CAG & CAG & NAG \\

2 &Indian govt is all Abt giving money to Bangladesh on the terms of Bangladesh ll give that all projects  to amabani n adani for thier benefits  lol who cares Abt soldiers or India they r just puppets of thier owners feku or pappu
 & CAG  & CAG & NAG\\

3 & When asked to speak in Parliament ran away. Speaks only in TV,radio or in election rally. Can we expect Another crying drama after Demonetisation disaster ? \#cryBaby"  & CAG &CAG &NAG\\ \bottomrule

\end{tabular}}
\label{tab:res-ana-cnn-cag}
\end{table}

\subsection{Significance Test}
To support our claim drawn in the previous section, significance tests, like Wilcoxon signed-rank test and Student t-test were carried out by comparing Weighted $F_1$ score of each classifier for each text representation scheme with fastText pre-trained vector scheme. Table \ref{tab:res:stats-en} and \ref{tab:res:stats-Hi} summarizes the p-values of statistical significance tests on English and Hindi Dataset respectively. In Wilcoxon signed-rank test, p-values of the results is less than 0.05 for Facebook English dataset and Twitter Hindi Dataset.  However, On Twitter English dataset and Facebook Hindi Dataset, some of the p-values are higher than 0.05. In student t-test, we get mixed bag results. By and large, our results are statistically significant.

\begin{table}
\tbl{p-values of Significance test on $F_1$-score on TRAC Facebook English Dataset}
{\begin{tabular}{ p{3cm}p{2cm}p{2cm} p{2cm} p{2cm} }\toprule

Text Rep. scheme &\multicolumn{2}{c}{Facebook English} &\multicolumn{2}{c}{Twitter English}\\ \cmidrule{2-5}

&Wilcoxon & T-test &Wilcoxon & T-test\\ \midrule

Count Vector &0.001 & \bf 0.12 &0.004&0.016\\

TF/IDF &0.0001& \bf 0.1465&0.0061 &0.0391\\

Word2vec &0.00002 &0.0001 & \bf 0.69 & \bf 0.30\\

Glove &0.00001 &0.000002 &0.01 &0.0009\\

fastText &0.00003 &0.0004& \bf 0.12 & \bf 0.08\\

doc2vec-dmc &0.000009 &0.0001 &0.0004 & 0.000004 \\

doc2vec-dbow &0.000009 &0.00003 &0.0004&0.000006 \\

p-Word2vec &0.00001 &0.0006 &0.02 &0.01 \\

P-Glove &0.0002 &0.02 &0.08& \bf 0.30 \\ \bottomrule

\end{tabular}}
%TODO : you should write a descriptive caption
%\caption{p-values of Significance test on $F_1$-score on TRAC English Dataset}
\label{tab:res:stats-en}
\end{table}

\begin{table}
\tbl{p-values of Significance test on $F_1$-score on TRAC code-mixed Hindi Test Dataset}
{\begin{tabular}{ p{3cm}p{2cm}p{2cm} p{2cm} p{2cm} }\toprule

Text Rep. scheme &\multicolumn{2}{c}{Facebook Code-mixed Hindi} &\multicolumn{2}{c}{Twitter Code-mixed Hindi}\\ \cmidrule{2-5}

&Wilcoxon & T-test &Wilcoxon & T-test\\ \midrule
Count Vector &0.003 & 0.05 &0.01& \bf 0.20\\

TF/IDF &0.003&\bf 0.14&0.007 & \bf 0.12\\

Word2vec & \bf 0.40 & \bf 0.17 &0.043 &0.017\\

Glove &0.001 &0.0005 &0.015 &0.004\\

fastText &\bf 0.35 &\bf 0.15&0.022 & 0.006\\

doc2vec-dmc &0.005 &0.00001 &0.001 & 0.0008 \\
doc2vec-dbow &0.003 &0.002 &0.001&0.002\\ \bottomrule

\end{tabular}}
%TODO : you should write a descriptive caption
%\caption{p-values of Significance test on $F_1$-score }
\label{tab:res:stats-Hi}
\end{table}

In the following subsection, we will try to answer all the research questions framed during the experiments were planned.

\subsection{ Best Text Representation scheme to model the text from Social web}
Text Representation is the primary task for to address any NLP task like Question/answering, classification etc. As dicusses in section \ref{Text-rep-scheme}, There are basically two text representing scheme:Bag-of-Word(BoW) with countvector, TF/IDF weighting and word embedding. Word2Vec \citep{mikolov2013efficient}, Glove \citep{pennington2014glove}, and fastText \citep{mikolov2018advances}, an extension of Word2vec are popular word embedding techniques. 

Results clearly show that models with fastText pre-trained vector outperform Glove pre-trained vector on Facebook test dataset as well as the Twitter test dataset. the main reason behind the outperformance of fastText over Glove and Word2vec is that The fastText consider each word as N-gram characters. A word vector for a word is computed from the sum of the n-gram characters. Glove and Word2vec consider each word as a single unit and provide a word vector for each word. Since Facebook users make a lot of mistakes in spelling, typos, fastText is more convenient than Glove \citep{majumder2018filtering}. from  Figure \ref{fig:hf} and \ref{fig:ht} shows that BoW is still effective text representation scheme for the standard machine learning classier which takes hand-crafted feature and n-grams as inputs. Logistic Regression and Support Vector perform better than other classifiers in English as well as Hindi Dataset. Adaboost performs better than LR and SVC on Facebook English Dataset but substantially underperform them on rest of three Datsets. Our participation \citep{majumder2018filtering}  in  TRAC competition \citep{Kumar18} FIRE Information Retrieval from Microblogs during Disasters \citep{fire2018-irdimis} track where our team performed well and secured top position.

\subsection{Transfer Learning Model vs Pre-trained Word Embedding Model}
Transfer learning is focused on storing knowledge gained while solving one problem and applying it to a different but related problem. On many occasion, NLP researchers face the problem of unavailability of sufficient labeled data to train the model. With the advent of new transfer learning method like ELMO \citep{peters2018deep} and Universal language model  fine-tuning for Text Classification (ULMFiT) \citep{howard2018universal} attract interest among NLP Researchers. These models are trained or large text corpus. \citet{howard2018universal} claimed that these model can be fine-tuned on the task-specific corpus.  We have used these transfer learning model on TRAC English dataset \cite{trac2018dataset} and results are presented in table \ref{tab:res:transfer}. one can observe that results are substantially lower than the results reported in Table  \ref{tab:res-en-fb-mc}, \ref{tab:res-en-tw-mc}, and \ref{tab:res-en-fbtw-dnn} where pre-trained word vectors are used to initialize the first layer of deep neural model and rest of the network is trained from scratch achieve better results than transfer learning model.  \citet{howard2018universal} termed use of pre-trained vector as shallow representation. 

%Transfer learning is essential for any kind of learning. Humans are not taught every single task or problem in order to be successful at it. Everyone gets into situations that have never been encountered, and we still manage to solve problems in an ad-hoc manner. The ability of learning from a large number of experiences, and exporting 'knowledge' into new environments is exactly what transfer learning is all about. Two common approaches  Develop Model Approach and Pre-trained Model Approach are used in this problem.

In these experiments, we trained different classifier models on Facebook posts. Table \ref{tab:res-en-tw-mc} \ref{tab:res-hi-tw-mc} shows the results on Twitter dataset \cite{trac2018dataset}. There is lexical difference between Facebook and Twitter posts. From the results shown in Table \ref{tab:res-en-tw-mc} \ref{tab:res-en-fbtw-dnn} and table \ref{tab:res-hi-tw-mc}, one can conclude that weighted $F_1$ score of standard machine learning classifiers are substantially lower in Twitter Dataset as compare to Facebook Dataset. While deep learning models perform better than machine learning classifiers for the Twitter Dataset. Thus, Deep learning models are more robust than machine learning classifier across diverse datasets.

\subsection{Does Deeper Neural Net make Sense}
To answer this question, we designed first CNN model with one convolution layer and other CNN model with 3 convolution layer with different filters height. As we look at results shown in Table \ref{tab:res-en-fb-mc},\ref{tab:res-en-tw-mc},\ref{tab:res-en-fbtw-dnn}, and \ref{tab:res-hi-fb-mc}, one can conclude that by and large weighted $F_1$ score lower for CNN model with multiple convolution layer than CNN model with single convolution layer. 

\section{Conclusion}
\label{lab:con}
In this Paper, Multilingual Social media stream is studied with special kind of text features: Aggression and fact perspective. Exhaustive experiments are performed to benchmark the text representation scheme on machine learning classifiers and deep neural nets. From the results, we conclude that deep Neural model with pre-trained  word embedding is the better choice than machine earning classifier and transfer learning model. Word embedding is the better text representative scheme than Bag-of-words for the deep neural models. In fact, performance can be improved with the help of fastText pre-trained vector. However, machine learning classifiers perform better in BoW with TF/IDF weighting than word embedding. We also concluded that higher drop out will help to counter model overfitting and improvise a standard evaluation metrics. CNN and LSTM  are the better models for these datasets. On the English test corpus, we obtained a better weighted $F_1$ score for NAG class and poor weighted $F_1$ score for CAG class which supports the previous \citep{malmasi2017detecting} findings. For the Facebook Hindi test corpus, the same seems not to be true. We obtained a better F1 score for CAG class than NAG class. It is also to be noted that the model leads to poor result on Twitter test data since the training corpus was created from Facebook. In such cases, deep neural models substantially outperform machine learning classifiers. Significance test confirms these claims with 95 \% confidence interval in most the cases. Our work shows what kind of problems are moving into the center of attention for research in machine learning. Using deep learning models, there is great potential to solve some of these problems, yet still, the performance is far from perfect. Model transfer between problems and the application of derived knowledge in user interfaces are areas directions for future work.


\begin{thebibliography}{}

\bibitem[Aroyehun et al.(2018)]{aroyehun2018aggression}
Aroyehun, Segun Taofeek and Gelbukh, Alexander (2018). \emph{Aggression
detection in social media: Using deep neural networks, data augmentation, and
pseudolabeling}.\emph{Proceedings of the First Workshop on Trolling, Aggression
and Cyberbullying (TRAC-2018)}, pp.90--97.

\bibitem[Arroyo et al. (2018)]{arroyo2018cyberbullying}
Arroyo-Fern{\'a}ndez, Ignacio and Forest, Dominic and Torres-Moreno,
Juan-Manuel and Carrasco-Ruiz, Mauricio and Legeleux, Thomas and Joannette,Karen
(2018). \emph{Cyberbullying Detection Task: the EBSI-LIA-UNAM System (ELU) at
COLING'18 TRAC-1}.\emph{Proceedings of the First Workshop on Trolling,Aggression
and Cyberbullying (TRAC-2018)}, pp 140--149.


\bibitem[Basu et al. (2018)]{fire2018-irdimis}
 Basu, Moumitaand Ghosh, Saptarshi and Ghosh, Kripabandhu (2018). \emph{Overview
 of the FIRE 2018  track: Information Retrieval from Microblogs during Disasters
 (IRMiDis)}. \emph{Proceedings of FIRE 2018 - Forum for Information Retrieval
 Evaluation}, Gujrat, India, December .


\bibitem[Baziotis et al. (2017)]{baziotis2017datastories}
Baziotis, Christos and Pelekis, Nikos and Doulkeridis, Christos (2017).
\emph{Datastories at semeval-2017 task 4: Deep lstm with attention for
message-level and topic-based sentiment analysis}. \emph{Proceedings of the
11th International Workshop on Semantic Evaluation (SemEval-2017)},
  pp.747--754

\bibitem[Bojanowski et al. (2017)]{bojanowski2017enriching}  
 Bojanowski, Piotr and Grave, Edouard and Joulin, Armand and Mikolov, Tomas
 (2017) \emph{Enriching word vectors with subword information},
 \emph{Transactions of the Association for Computational Linguistics},
  vol-5,pp.{135--146}, MIT Press.

\bibitem[Burnap et al. (2015)]{burnap2015cyber}
 Burnap, Pete and Williams, Matthew Ltitle (2015). \emph{Cyber hate speech on
 twitter: An application of machine classification and statistical modeling for
 policy and decision making}\emph{Policy \& Internet}, vol-7 number 2, 
 pp.223--242,  Wiley Online Library.

\bibitem[Conover et al. (2011)]{conover2011political}
Conover, Michael and Ratkiewicz, Jacob and Francisco, Matthew R and
Gon{\c{c}}alves   , Bruno and Menczer, Filippo and Flammini, Alessandro (2011).
\emph{Political polarization on twitter.},\emph{Icwsm},  vol-133, pp.89--96.   

\bibitem[Conover Micheal et al. (2011)]{conover2011predicting}
Conover, Michael D and Gon{\c{c}}alves, Bruno and Ratkiewicz, Jacob and
Flammini, Alessandro and Menczer, Filippo (2011).  \emph{Predicting the
political alignment of twitter users},  \emph{Privacy, Security, Risk and Trust
(PASSAT) and 2011 IEEE Third Inernational Conference on Social Computing
(SocialCom), 2011 IEEE Third International Conference on},
  pp.192--199.

\bibitem[Davidson et al. (2017)]{davidson2017automated}
Davidson, Thomas and Warmsley, Dana and Macy, Michael and Weber, Ingmar
(2017). \emph{Automated Hate Speech Detection and the Problem of Offensive
Langemuage}. \emph{Proceedings of ICWSM}.

\bibitem[Deriu et al. (2016)]{deriu2016swisscheese}
Deriu, Jan and Gonzenbach, Maurice and Uzdilli, Fatih and Lucchi, Aurelien and
Luca, Valeria De and Jaggi, Martin (2016). \emph{Swisscheese at semeval-2016
task 4: Sentiment classification using an ensemble of convolutional neural
networks with distant supervision}. \emph{Proceedings of the 10th international
workshop on semantic evaluation},pp.1124--1128.


\bibitem[Hltcoe et.al (2013)]{hltcoe2013semeval}
Hltcoe, J (2013).\emph{Semeval-2013 task 2: Sentiment analysis in
Twitter},vol-312  Atlanta, Georgia, USA.
  

\bibitem[Howaard et al.(2018)]{howard2018universal}
Howard, Jeremy \& Ruder, Sebastian {2018}. \emph{Universal language model
fine-tuning for text classification"},\emph{arXiv preprint arXiv:1801.06146},
  
\bibitem[Harris et al. (1954)]{harris1954distributional}
Harris, Zellig S (1954),\emph{Distributional structure} Word,{10}, 
number={2-3},pp.146--162, \emph{1954}, Taylor \& Francis.

\bibitem[Kumar et al. (2018)]{trac2018dataset}
Kumar, Ritesh and Reganti, Aishwarya N. and Bhatia, Akshit and Maheshwari,Tushar
(2018), \emph{Aggression-annotated Corpus of Hindi-English Code-mixed Data},
\emph{Proceedings of the 11th Language Resources and Evaluation Conference
(LREC)}, Miyazaki, Japan.

\bibitem[Kumar Ritesh et al.(2018)]{Kumar18}
Kumar, Ritesh and Ojha, Atul Kr. and Malmasi, Shervin and Zampieri Marcos (2018). \emph{{Benchmarking Aggression Identification in Social Media},
\emph{Proceedings of the First Workshop on Trolling, Aggression and Cyberbulling (TRAC)}},  Santa Fe, USA 


\bibitem[Kwok et al. (2013]{kwok2013locate}
 Kwok, Irene and Wang, Yuzhou (2013),\email{Locate the hate: Detecting Tweets
 Against Blacks},\emph{Twenty-Seventh AAAI Conference on Artificial
 Intelligence}.

\bibitem[Lau et al. (2016)]{lau2016empirical}
Lau, Jey Han and Baldwin, Timothy (2016). \emph{An empirical evaluation of
doc2vec with practical insights into document embedding generation}. \emph{arXiv
preprint arXiv:1607.05368}.


\bibitem[Le, Quoc et al. (2014)]{le2014distributed}
Le, Quoc and Mikolov, Tomas (2014) \emph{Distributed representations of
sentences and documents}.\emph{International Conference on Machine Learning},
pp.1188--1196 

\bibitem[Majumder et al.(2018)]{majumder2018filtering}
Majumder, Prasenjit and Mandl, Thomas and Modha Sandip (2018)\emph{Filtering
Aggression from the Multilingual Social Media Feed} \emph{Proceedings of the
First Workshop on Trolling, Aggression and Cyberbullying (TRAC-2018)}, pp.
{199--207}


\bibitem[Malmasi, et al. (2017)]{malmasi2017detecting}
Malmasi, Shervin and Zampieri, Marcos (2017)\emph{Detecting Hate Speech in
Social Media} \emph{Proceedings of the International Conference Recent Advances
in Natural Language Processing (RANLP)},  pp.{467--472}.

\bibitem[Malmasi et al. (2018)]{malmasi2018challenges}
Malmasi, Shervin and Zampieri, Marcos (2018).\emph{Challenges in Discriminating
Profanity from Hate Speech} \emph{Journal of Experimental \& Theoretical
Artificial Intelligence}  pp.{1--16},  vol-30,  issue-2,Taylor \& Francis.


\bibitem[Maynard et al.(2011)]{maynard2011automatic}
Maynard, Diana and Funk, Adam (2011) \emph{Automatic detection of political
opinionsin tweets}, \emph{Extended Semantic Web Conference},pp. {88--99},
\emph{Springer}.
  
\bibitem[Mikolov et.al (2013)]{mikolov2013efficient}
Mikolov, Tomas and Chen, Kai and Corrado, Greg and Dean, Jeffrey (2013).
\emph{Efficient estimation of word representations in vector space},\emph{arXiv
preprint arXiv:1301.3781}.



 
\bibitem[Mikolov et al. (2018)]{mikolov2018advances}
 Mikolov, Tomas and Grave, Edouard and Bojanowski, Piotr and Puhrsch, Christian
 and Joulin, Armand (2018). \emph{Advances in Pre-Training Distributed Word
 Representations},\emph{Proceedings of the International Conference on Language
 Resources and Evaluation (LREC 2018)}.

\bibitem[Modha et al. (2016)]{modha2016daiict}
Modha, Sandip and Agrawal, Krati and Verma, Deepali and Majumder, Prasenjit and
Mandalia, Chintak 2016. \emph{DAIICT at TREC RTS 2016: Live Push Notification
and Email Digest.},TREC.


\bibitem[Mohammad et al.(2013)]{mohammad2013nrc}
Mohammad, Saif M and Kiritchenko, Svetlana and Zhu, Xiaodan (2013).
\emph{NRC-Canada: Building the state-of-the-art in sentiment analysis of
tweets.}\emph{arXiv preprint arXiv:1308.6242}.

\bibitem[Nakov et al. (2016)]{nakov2016semeval}
Nakov, Preslav and Ritter, Alan and Rosenthal, Sara and Sebastiani, Fabrizio and
Stoyanov, Veselin (2016). \emph{SemEval-2016 task 4: Sentiment analysis in
Twitter},\emph{Proceedings of the 10th international workshop on semantic
evaluation (semeval-2016)},  pp. {1--18}.

\bibitem[Pennington et.al. (2014)]{pennington2014glove}
Pennington, Jeffrey and Socher, Richard and Manning, Christopher \emph{Glove:
Global vectors for word representation} \emph{Proceedings of the 2014 conference
on empirical methods in natural language processing (EMNLP)},
pp.{1532--1543}.


\bibitem[Peters et al. (2018)]{peters2018deep}
Peters, Matthew E and Neumann, Mark and Iyyer, Mohit and Gardner, Matt and
Clark, Christopher and Lee, Kenton and Zettlemoyer, Luke (2018). \emph{Deep
contextualized word representations} \emph{arXiv preprint arXiv:1802.05365}.



\bibitem[Razavi et.al (2010)]{razavi2010offensive}
Razavi, Amir H and Inkpen, Diana and Uritsky, Sasha and Matwin, Stan (2010)
\emph{Offensive language detection using multi-level classification},
  \emph{Canadian Conference on Artificial Intelligence} pp.{16--27},
  Springer


\bibitem[Rosenthal et al. (2015)]{rosenthal2015semeval}
Rosenthal, Sara and Nakov, Preslav and Kiritchenko, Svetlana and Mohammad, Saif
and Ritter, Alan and Stoyanov, Veselin (2015). \emph{Semeval-2015 task 10:
Sentiment analysis in twitter} \emph{Proceedings of the 9th international
workshop on semantic evaluation (SemEval 2015)},   pp.{451--463}.
 

\bibitem[Rosenthal et al. (2017)]{rosenthal2017semeval}
Rosenthal, Sara and Farra, Noura and Nakov, Preslav (2017). \emph{SemEval-2017
task 4: Sentiment analysis in Twitter} \emph{Proceedings of the 11th
International Workshop on Semantic Evaluation (SemEval-2017)}, pp.{502--518}.



\bibitem[Schmidt et al. (2017)]{schmidt2017survey}
Schmidt, Anna and Wiegand, Michael (2017).\emph{A Survey on Hate Speech
Detection Using Natural Language Processing},\emph{Proceedings of the Fifth
International Workshop on Natural Language Processing for Social Media.
Association for Computational Linguistics} {Valencia, Spain}, pp.{1--10},


\bibitem[Severyn et al. (2015)]{severyn2015unitn}
Severyn, Aliaksei and Moschitti, Alessandro (2015) \emph{Unitn: Training deep
convolutional neural network for twitter sentiment classification},
\emph{Proceedings of the 9th international workshop on semantic evaluation
(SemEval 2015)}, pp. {464--469}.

\bibitem[Tumasjan et al. (2010)]{tumasjan2010predicting}
Tumasjan, Andranik and Sprenger, Timm Oliver and Sandner, Philipp G and Welpe,
Isabell M (2010). \emph{Predicting elections with twitter: What 140 characters
reveal about political sentiment.},\emph{Icwsm},  vol-10,  number-{1},
  pp. {178--185}.

\bibitem[Warner et al. (2012)]{warner2012detecting}
Warner, William and Hirschberg, Julia (2012) \emph{Detecting hate speech on the world wide web}, \emph{Proceedings of the Second Workshop on Language in Social Media},pp {19--26},{Association for Computational Linguistics}.


\bibitem[xu et al. (2012)]{xu2012learning}
 Xu, Jun-Ming and Jun, Kwang-Sung and Zhu, Xiaojin and Bellmore, Amy (2012).
 \emph{Learning from bullying traces in social media}, \emph{Proceedings of the
 2012 conference of the North American chapter of the association for
 computational linguistics: Human language technologies}, pp.{656--666},
{Association for Computational Linguistics}
  
\bibitem[Zhang et al. (2015)]{zhang2015sensitivity} 
  Zhang, Ye and Wallace, Byron (2015). \emph{A Sensitivity Analysis of (and Practitioners' Guide to) Convolutional Neural Networks for Sentence Classification},\emph{arXiv preprint arXiv:1510.03820}.
\bibitem[Majumder et al. (2008)]{majumder2008text}
 Majumder, Prasenjit and Mitra, Mandar and Pal, Dipasree and Bandyopadhyay, Ayan and Maiti, Samaresh and Mitra, Sukanya and Sen, Aparajita and Pal, Sukomal.\emph {Text collections for FIRE},
 \emph{Proceedings of the 31st annual international ACM SIGIR conference on research and development in information retrieval},pp.{699--700},{ACM}


\bibitem[Majumder et al. (2007)]{majumdar2007initiative}
Majumdar, P and Mitra, Mandar and Parui, Swapan K and Bhattacharya, \emph{Initiative for indian language ir evaluation}, \emph{The First International Workshop on Evaluating Information Access (EVIA)}


\end{thebibliography}
\end{document}